\documentclass[aps,prl,twocolumn,superscriptaddress,amsmath,amssymb,showpacs]{revtex4}

\usepackage{graphicx}
\usepackage{amsmath}
\begin{document}

\title{Efficient ion acceleration by collective laser-driven electron dynamics with ultra-thin foil targets}

\author{S. Steinke}
\email[]{steinke@mbi-berlin.de}
\affiliation{Max-Born-Institut, D-12489 Berlin, Germany}

\author{A. Henig}
\affiliation{Max-Planck-Institut f.\ Quantenoptik, D-85748 Garching, Germany}
\affiliation{Fakult\"at f.\ Physik, LMU M\"unchen, D-85748 Garching, Germany}

\author{M. Schn\"urer}
\author{T. Sokollik}
\author{P.\,V. Nickles}
\affiliation{Max-Born-Institut, D-12489 Berlin, Germany}

\author{D. Jung}
\author{D. Kiefer}
\affiliation{Max-Planck-Institut f.\ Quantenoptik, D-85748 Garching, Germany}
\affiliation{Fakult\"at f.\ Physik, LMU M\"unchen, D-85748 Garching, Germany}

\author{J. Schreiber}
\affiliation{Max-Planck-Institut f.\ Quantenoptik, D-85748 Garching, Germany}
\affiliation{Fakult\"at f.\ Physik, LMU M\"unchen, D-85748 Garching, Germany}
\affiliation{Imperial College London, SW7 2BZ, UK}

\author{T. Tajima}
\affiliation{Fakult\"at f.\ Physik, LMU M\"unchen, D-85748 Garching, Germany}
\affiliation{Photomedical Research Center, JAEA. Kyoto, Japan}

\author{X.\,Q. Yan}
\affiliation{Max-Planck-Institut f.\ Quantenoptik, D-85748 Garching,
Germany} \affiliation{State Key Lab of Nuclear physics and
technology, Peking University, 100871, Beijing, China}

\author{J. Meyer-ter-Vehn}
\affiliation{Max-Planck-Institut f.\ Quantenoptik, D-85748 Garching,
Germany}

\author{M. Hegelich}
\affiliation{Fakult\"at f.\ Physik, LMU M\"unchen, D-85748 Garching, Germany}
\affiliation{Los Alamos National Laboratory, Los Alamos, New Mexico 87545, USA}

\author{W. Sandner}
\affiliation{Max-Born-Institut, D-12489 Berlin, Germany}

\author{D. Habs}
\affiliation{Fakult\"at f.\ Physik, LMU M\"unchen, D-85748 Garching, Germany}
\affiliation{Max-Planck-Institut f.\ Quantenoptik, D-85748 Garching, Germany}

\begin{abstract}
Experiments on ion acceleration by irradiation of ultra-thin diamond-like carbon (DLC) foils, with thicknesses well below the skin depth, irradiated with laser pulses of ultra-high contrast and linear polarization, are presented. A maximum energy of 13\,MeV for protons and 71\,MeV for carbon ions is observed with a conversion efficiency of $\sim 10\%$. Two-dimensional particle-in-cell (PIC) simulations reveal that the increase in ion energies can be attributed to a dominantly collective rather than thermal motion of the foil electrons, when the target becomes transparent for the incident laser pulse.
\end{abstract}


\maketitle

Recent experiments in the field of relativistic laser-plasma interaction have
shown that the conversion efficiency (CE) from the laser to the kinetic energy of
the ion bunch  as well as the maximal energy of the ions can be improved by
the use of ultra-thin foil targets in the range of 100~nm
\cite{steinke2009-01, Neely2006-01, Ceccotti2007-01, antici2007-01}.
Additionally, it was emphasized that
besides the standard target normal sheath acceleration (TNSA) \cite{Hatchett1999-01, wilks1992-01} other,
radiation pressure dominated acceleration mechanisms become possible for
ultra-thin targets \cite{klimo2008-01, yan2008-01, robinson2008-01}.
An ultra-high contrast is required to avoid substantial expansion of the targets before
the interaction with the main pulse. These conditions can be achieved by the use of a specially designed laser system in combination with double plasma
mirrors (DPM) \cite{wittmann2006-01, steinke2009-01, levy2007-01}.
The partial transmission of the intense laser pulse
through an expanding target is expected to play a decisive role \cite{Humieres2005-01,yin2006-01, Albright2007-01} and has been lately demonstrated in the regime of relativistic transparency \cite{Henig2009}.

In this letter we present experimental results on ion acceleration from DLC foils of thicknesses ranging from 50\,nm down to 2.9\,nm. The targets are irradiated by linear polarized pulses of 45\,fs FWHM duration focussed to a peak intensity of up to $5\times10^{19}\,$W/cm$^2$. We find an optimum in ion acceleration at a target thickness of 5.6\,nm, where ion energies reach values of 13\,MeV for protons and 71\,MeV for carbon ions. Thus, for the ultra-short pulses used, the optimum foil thickness is well below the collisional skin depth $l_s\sim 13\,$nm of the heated target, which is thus becoming transparent to the laser. The corresponding CE is reaching values of $\sim 1.6$\% for protons ($E>2MeV$) and $\sim 10$\% in the case of C$^{6+}$ ($E>5MeV$).
By means of 2D PIC-simulations we find that the transmitted laser field imposes a dominantly collective rather than thermal motion of the foil electrons, leading to increased ion energies and enhanced CE. Our findings are supported by a semi-analytical model, which shows good agreement with the experimental results.

The experiments were performed at the MBI - TW Ti:sapph laser of central wavelength 810\,nm delivering 1.2\,J in
45\,fs FWHM pulses with an amplified spontaneous emission (ASE) contrast ratio smaller than $10^{-7}$ up to $\sim10\,$ps prior to the arrival of the main peak. By means of a re-collimating DPM
\cite{steinke2009-01}, this contrast was increased by estimated four orders of
magnitude \cite{levy2007-01}, which is essential for the suppression of pre-heating and expansion due to the pulse background.
The energy throughput of this DPM system was improved to
values of 60-65\,\% compared to our previous experiments \cite{steinke2009-01}, resulting in pulse energies of 0.7\,J. Finally, the laser pulse
 was focused on the DLC target with a f/2.5 parabolic mirror down to 6
$\mu$m diameter and is diffraction limited by 30\,\% under normal incidence.
This corresponds to a peak intensity of $~2.6\ \times 10^{19}$ W/cm$^2$ or
a normalized vector potential of $a_L=3.6$. For a second set of experiments
the focussing procedure was improved, leading to a focus diameter of 3.6$\,\mu$m
and therefore a peak intensity of $5\ \times 10^{19}$ W/cm$^2$ or $a_L=5$.
The resulting ions were registered with a MCP coupled to a
Thomson-Parabola \cite{sargis2005-01}.

DLC targets of thicknesses ranging from $2.9-50\,$nm were used, having a density of 2.7 g/cm$^3$. Owing to the high fraction of sp$^3$-, i.e. diamond-like bonds of $\sim$75\%, DLC offers unique properties for the production of mechanically stable, ultra-thin, free standing targets, such as exceptionally high tensile strength, hardness and heat resistance. The thickness of the DLC foils was characterized by means of an atomic force microscope (AFM), including the hydrocarbon contamination layer on the target surface which was present during the experiments. In addition, in order to precisely determine the structure of the contamination layer, the depth-dependend composition of the foil was measured via Elastic Recoil Detection Analysis (ERDA). From these measurements we obtain a thickness of $\sim 1\,$nm for the hydrocarbon contamination layer. Throughout the manuscript we are referring to the combined thickness of bulk and surface layer as it appears in the actual ion acceleration experiment presented.

\begin{figure}[!t]
\centerline{\includegraphics[angle=0,width=1\columnwidth]
{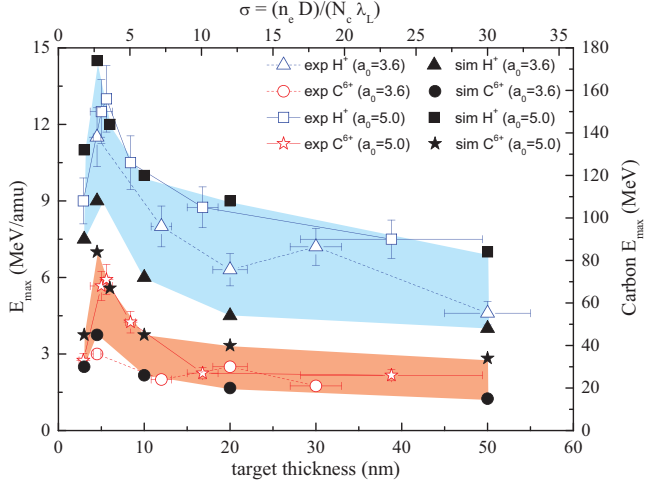}}
\caption{Maximum energy per atomic mass unit for both protons and carbon ions plotted vs. foil thickness for values of the normalized laser vector potential of $a_0=3.6$ and $a_0=5$. The experimental results are in excellent agreement with numbers deduced from 2D-PIC-simulations. }
\label{fig1}
\end{figure}

The maximum detectable energy/nucleon (\emph{E$_\textnormal{max}$}) obtained for protons and fully ionized carbon ions C$^{6+}$ is
plotted as a function of the target thickness in
FIG. \ref{fig1} for two different laser intensities. The cut-off energies for both ion species exhibit a strong dependence on target thickness. In case of protons and $a_0=5$, \emph{E$_\textnormal{max}$} increases from around 7.5\,MeV for a 40\,nm foil up to 13\,MeV for a 5.6\,nm foil, while for C$^{6+}$ the maximum energy rises from 26\,MeV for use of a 40\,nm foils to 71\,MeV for 5.6\,nm. Further decrease of the target thickness down to 2.9\,nm results in a steep drop of the observed ion energies.
The energy distributions of all species are continuous as expected. However, especially the observed carbon C$^{6+}$ energy of 71\,MeV reaches for the first time a range of values that were previously only accessible by large single shot Nd:glass laser systems with $30 - 50\,$J pulse energy \cite{Henig2009, Krushelnick2005-01}. The shot-to-shot variations of 10\,\%, indicated by the error bars, arise mainly from fluctuations of the laser pulse itself and from macroscopic modulations of the target surface.

\begin{figure}[!t]
\centerline{\includegraphics[width=1\columnwidth]{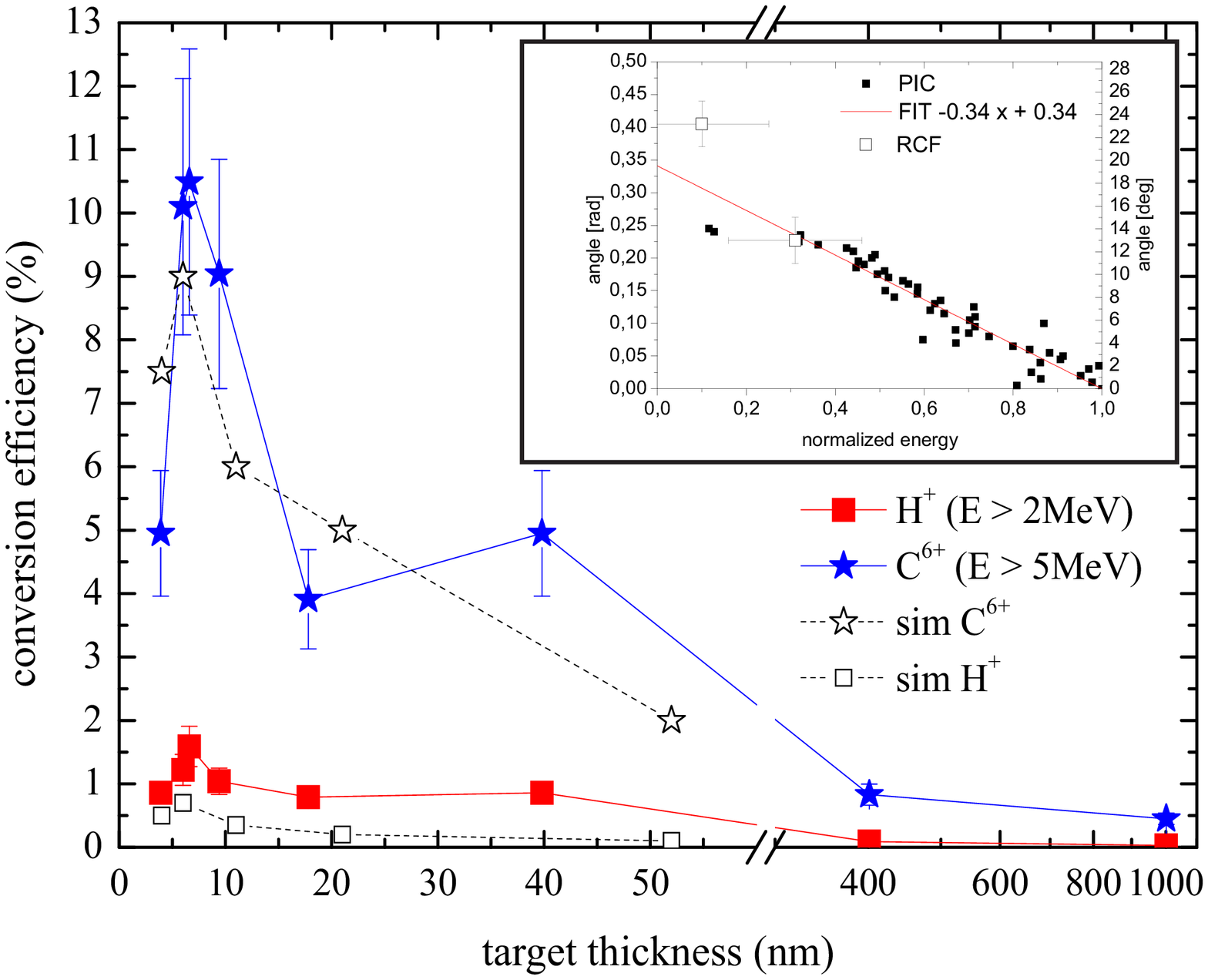}}
\caption{Calculated CE for protons and C$^{6+}$ ions as a function of the target thickness, based on a energy dependent divergence angle of the ion beam, which was extracted from PIC simulations and supported by measured beam profiles (cf. figure inset). The experimental data are in good agreement with the values of the CE obtained from PIC simulations.} \label{fig3}
\end{figure}

Furthermore, the CE (FIG.\ref{fig3}) was calculated by numerically convoluting an energy dependent divergence of the accelerated ion beam with the initially measured ($a_0=5$) spectra. These values (cf. inset in FIG.\ref{fig3}) were extracted from PIC simulations and supported by experimentally obtained proton beam profiles using a stack of radiochromic film (RCF) layers. Comparing those to the divergencies of ion beams generated by other laser systems e.g. \cite{nuernberg2009}, our divergence is smaller and the linear dependence on the energy is reasonable in the considered high-energy part.
The dependence of the CE on the target thickness is showing analogous characteristics to \emph{E$_\textnormal{max}$} (cf. FIG.1). A sudden rise is observed when entering the laser transparency regime ($D < l_s$) reaching 10.5\,\% for $C^{6+}$ and 1.6\,\% for protons at the optimum thickness.
The CE drops down to below 1\% when the target thickness reaches the TNSA dominated regime \cite{Neely2006-01, fuchs2006}.

The experiments were compared to 2D-PIC simulations, where the laser pulse was modeled by a Gaussian shape in time with a FWHM of 16 laser cycles, a Gaussian intensity distribution in focus with a FWHM spot size of 4~$\mu$m and an $a_0=5$ ($a_0=3.6$). It interacts with a rectangular
shaped plasma of initial density $n_e=500 \times n_c$
consisting of 90\,\% carbon ions and 10\,\% protons (in number density) to account for
the presence of a contamination layer.
The simulation box is composed of 1200 $\times$ 10000 cells with 1000 particles per cell and a total size of (10 $\times$ 20) $\mu$m$^2$. The total simulation time $\tau$ given in laser cycles is 120.

FIG.~1 shows that the simulated maximum proton energies as well as the carbon
energies are in excellent agreement with the
experimental data. In particular, the optimum target thickness of 5.6\,nm
is reproduced. This thickness for the peak ion energy is consistent with the
thickness given by the empirical relation
$
\sigma \stackrel{!}{\approx} 3+0.4\times a_0=5
$
that was found in multiparametric PIC-simulation studies by Esirkepov \emph{et al.} \cite{Esirkepov2006}.
Here, a normalized areal electron density $\sigma= (n_e/n_c)\cdot(D/\lambda_L)$
is introduced, where $D$ is the target thickness, $\lambda_L$ the
laser wave length, $n_e$ the electron density of the target and
$n_c$ the critical electron density, which in our case calculates to $\sigma = 4.6$. For foil thicknesses below the optimum ($\sigma < 3+0.4\times a_0$), the plasma becomes increasingly transparent and the pulse is more transmitted than absorbed.
Due to the low number of electrons in the focal volume ($\sim 10^{11}$) their electric current is no longer sufficient to (i) reflect the laser pulse and (ii) to establish an effective longitudinal charge separation field. This results
in a sudden drop in ion energies and $50\,\%$ of the CE (the same amount as the reduction of the target thickness), as it was observed in the experiment (see FIG.~1+~2).
 In case of $\sigma > 3+0.4\times a_0$, the laser intensity is not sufficient to generate the maximum possible displacement of all electrons within the focal volume which gives rise to a decrease in the longitudinal charge separation field.  Note that the optimum
thickness for ion acceleration predicted theoretically and observed experimentally is much smaller than
previously used target thicknesses.

We compared our results to a number of analytical scaling laws that have been derived for the standard scenario of ion acceleration from the back surface of a thin, but opaque foil, i.e. target normal sheath acceleration (TNSA) \cite{fuchs2006, schreiber2006-01, Andreev2008-01}. However, we find that all of the existing analytical models fail to even qualitatively predict the two main features seen experimentally, namely the distinct peak in ion energies at $\sigma \approx 3+0.4\times a_0$ as well as the strong dependence of the cut-off value on target thickness. Consequently, a refined theory is needed.

\begin{figure}[!ht]
\centerline{\includegraphics[width=1\columnwidth]
{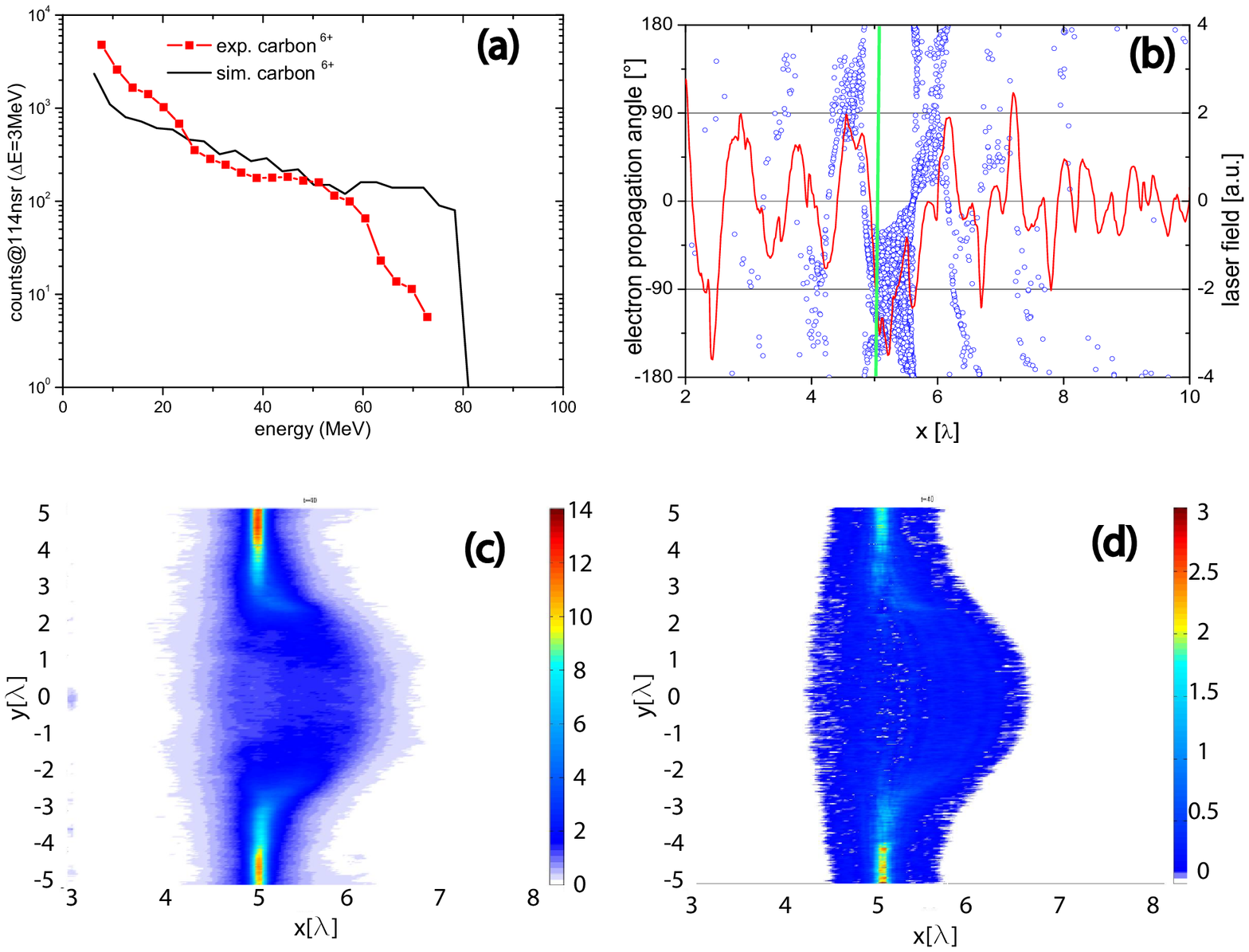}}
\caption{(a) Comparison between experimentally obtained and simulated $C^{6+}$ spectrum as deduced from a 2D-PIC simulation of a $4.5\,$nm foil target (b) Electron propagation angle (blue circles) and laser transverse electric field $E_y$ (red line) at time t$=30\,\tau$, plotted vs. laser propagation direction $x$, given in units of the laser wavelength $\lambda$. $\tau$ is given in laser cycles. The green line at $x=5\lambda$ gives the initial location of the target. For the same simulation, (c) gives the electron and (d) the carbon density distribution vs. $x$ at time t$=40\tau$.}
\label{fig2}
\end{figure}

In FIG. 3a the observed carbon energy spectrum is shown along with the simulated one agreeing with each other in their overall spectral shape.
To illustrate the underlying ion acceleration mechanism in our case we show in FIG.~\ref{fig2}c,d
detailed electron and carbon densities after 40 laser
cycles as a function of the laser propagation direction $x$ and the
transverse coordinate $y$ in units of the laser wave length
$\lambda_L$. It is immediately striking that ions are accelerated strongly asymmetric heavily favoring the direction of laser propagation.
This is in strong contrast to the model based on the self-consistent solution of the Poisson equation presented in \cite{Andreev2008-01} which predicts a symmetric acceleration that happens primarily after the end of the laser pulse. From FIG. 3c+d it can be seen, that the carbon ions are accompanied by co-moving electrons which accelerate the ions in forward direction.

Unlike the TNSA consideration in case of thicker targets, in our current case, coherence of laser-driven electron motion manifests importantly. In the typical TNSA with thicker targets, none of the laser field is transmitted as the foil remains opaque throughout the interaction duration. The electron motion is an indirect result of interaction with the laser at the front surface of the target and subsequent multiple interactions with matter in the target bulk. That scenario leads to thermal motion as the electrons traverse the target to set up a quasi-static field at the back. In contrast, in nm-scale foils of thickness below the skin depth the laser imprints a directed motion onto electrons as those targets are partially transparent. These coherent electron motions can not be treated as thermal motion, but rather as collective motion tied to the laser and thus directly to the laser amplitude a$_0$.
The propagation angle of electrons upon laser interaction with the target as extracted from the 2D PIC simulation at time $30\,\tau$ is shown in FIG. 3b. Obviously, the electron dynamics are classified in three categories.
(i) electrons are driven forward by laser,
(ii) electrons are forced to return backwards due to the electrostatic field,
(iii) electrons are performing a collective, transversal oscillation owing to the laser-created coherent wave structure.
Additionally, the laser transmission coefficient can be determined to be $T\sim$ 0.1, which agrees with a theoretical value of $T \cong 1/[1+(\pi\sigma/\gamma_p)^2]$, where  $\gamma_p$ is the Lorentz factor in the laser
$\sqrt{a_L^2+1}$ \cite{Vshivkov1998-01}.

Initially, without assuming thermalized electrons (i.e. in contrast to \cite{mora2003}),
we now formulate the maximal ion energies
obtained in the laser driven foil interaction in a semi-analytical model based on \cite{Mako1984}.

Following the analysis of
Mako \& Tajima \cite{Mako1984} and Yan et al. \cite{yan2009}, the forward current
density of electrons $J$ and electron density $n_e$ are related through
$ J = -e\int_{v}^{V_{\text{max}}}V_x g \text{d}V_x$
and
$n_e = \frac{2}{e}\int_{0}^{V_{\text{max}}}\frac{\text{d}J/\text{d}v}{v}
\text{d}v$,
where $g$ is the electron distribution function and
$V_{\text{max}} = c\sqrt{1-m_e^2 c^4/(E_{0}+e\phi)^2}$.
Here we note that at a given position the electron energy $E_{0}$ in the
electrostatic potential $\phi$ is given as an
implicit function of space and time through
$\varepsilon = \gamma m_ec^2 -e\phi$.
We find that
$
J(\varepsilon) = -J_0\left(1-\varepsilon/E_{0}\right)^\alpha
$
where $\alpha$, a measure for the coherence is found to be about 3 in our simulation for a target thickness of 4.5\,nm.
Considering the relativistic dynamics of electrons in our regime, the
electron density as a function of the current density may be integrated to
yield
\begin{align}
n_e &=
\frac{2}{e}\int_{0}^{V_{\text{max}}}\frac{\text{d}J/\text{d}v}{v}\text{d}v
=
\frac{2}{ec}\int_{-e\phi}^{E_{\text{max}}}\frac{\text{d}J}
{\text{d}\varepsilon}\text{d}\varepsilon
\notag
\\
&= \frac{2J_0}{ec}\left(1+\frac{e\phi}{E_0}\right)^\alpha.
\label{eq2}
\end{align}
The integration of the nonlinear coupled equations for nonrelativistic ions
under the self-consistently evolving potential using Eq.~\ref{eq2}
with the boundary
condition and initial value settings of Mako and Tajima leads to the maximum
ion (with charge $Q$) energy as
$
\varepsilon_{\text{max}} = (2\alpha +1) Q E_{0}
\label{eq3}
$, i.\,e.\ the maximum carbon energy is determined by the maximum electron
kinetic energy.
In the present experimental condition we recognize that the electron kinetic
energy is primarily dominated by the laser driven collective momentum so that
\begin{equation}
\varepsilon_{\text{max}} =(2\alpha+1)Qmc^2(\gamma_p-1)
\label{eq4}
\end{equation}
which gives rise to
$(2\alpha+1)Qmc^2\left(\sqrt{a_L^2+1}-1\right)$.

When we insert the
exponent of $\alpha=3$
as observed in our 2D-PIC simulations, we obtain for the case of
$a_0=3.6$
a theoretical maximum carbon (proton) energy of 59~MeV (9.6~MeV), in the case of $a_0=5$, 88~MeV  (14.3~MeV).
This is in agreement with the experimental value between 36-71 ~MeV (11-13~MeV).

Again, for conventional TNSA we do not expect a strong dependence of
the maximum ion energy on target thickness when the target
is much smaller than the transversal size of the laser focal
spot as it is the case in the present study. The thickness
dependence in TNSA is mainly due to transversal spread
of the laser accelerated electrons on their way through
the target \cite{schreiber2006-01, nickles2007-01}.

Instead, the strong increase in $\varepsilon_{\text{max}}$ as it was observed
in the experiment is related to the
growing transmission for thinner targets leading to collective electron dynamics imposed by the laser field.
For thick targets, however, $\alpha$ drops towards zero and Eq. 2 reveals the dependence of $\varepsilon_{\text{max}}$ on the hot
electron energy $mc^2(\sqrt{a_L^2+1}-1)$
\cite{mora2003,wilks1992-01}.

In summary, ion acceleration from ultra-thin DLC foils of thickness $2.9-50\,$nm was studied. A strong dependence of the resulting maximum energies on target thickness was observed experimentally, with a pronounced optimum for an initial foil extension of $5.6\,$nm with a very high conversion efficiency from the laser to the kinetic energy of the ion bunch. Here, proton energies of 13\,MeV with a CE of $1.6\, \%$ and carbon ions C$^{6+}$ of 71\,MeV with a CE of $10\, \%$ were obtained with laser pulse energies as low as $0.7\,$J. Our experimental results are in excellent agreement with 2D-PIC simulations. However, we find that previously published TNSA scaling laws based on analytical models \cite{fuchs2006, schreiber2006-01, Andreev2008-01} fail to interpret our results. Therefore, a new semi-analytical model based on \cite{Mako1984} was developed, showing good agreement with the experimental data.

Demonstrating enhanced ion energies and CE from table-top laser systems operating at high repetition rates represents a major step towards potential applications such as medical or industrial motivated approaches.

\begin{acknowledgments}
We thank R. Hoerlein for fruitful discussions.
This work was partly supported by Deutsche Forschungsgemeinschaft
through Transregio SFB TR18 and the DFG-Cluster of Excellence Munich-Centre for Advanced
Photonics (MAP). A.~Henig, D.~Kiefer and D.~Jung acknowledge financial support
from IMPRS-APS, J.~Schreiber from DAAD, X.~Q.~Yan from the
Humboldt foundation and NSFC(10855001).
\end{acknowledgments}


\begin{thebibliography}{99}

\bibitem{steinke2009-01}A.\,A.~Andreev et al. , Physics of Plasmas {\bf
16}, 013103 (2009).

\bibitem{Neely2006-01}D.~Neely et al., Appl.\ Phys.\ Lett. {\bf 89},
021502 (2006).

\bibitem{Ceccotti2007-01}T.~Ceccotti et al., Phys.\ Rev.\ Lett. {\bf 99},
185002 (2007).

\bibitem{antici2007-01}A.\,A.~Antici et al. , Physics of Plasmas {\bf
14}, 030701 (2007).

\bibitem{Hatchett1999-01}S.\,P.~Hatchett et al., Physics of Plasmas {\bf
7}, 2076 (2000).

\bibitem{wilks1992-01}S.\,C.~Wilks et al., Phys.\ Rev.\ Lett. {\bf 69}, 1383
(1992).

\bibitem{yan2008-01}X.\,Q.~Yan et al., Phys.\ Rev.\ Lett. {\bf 100}, 135003
(2008).

\bibitem{robinson2008-01}A.\,P.\,L.~Robinson et al., New J.\ Phys. {\bf 10}, 13 (2008).

\bibitem{klimo2008-01} O.~Klimo et al., Phys.\ Rev.\ ST AB {\bf 11}, 031301 (2008).


\bibitem{wittmann2006-01} T.~Wittmann et al., Review of Scientific Instruments {\bf{77}}, 083109 (2006).

\bibitem{levy2007-01} A.~L\'{e}vy et al., Opt. Lett. {\bf{32}}, 310 (2007).

\bibitem{Humieres2005-01} E.~d'Humieres et al., Physics of Plasmas {\bf{12}}, 062704 (2005).

\bibitem{yin2006-01}L.~Yin et al., Laser and Particle Beams {\bf{24}}, 291 (2006).

\bibitem{Albright2007-01} B. J.~Albright et al., Physics of Plasmas {\bf{14}}, 094502 (2007).

\bibitem{Henig2009} A.~Henig et al., Phys. Rev. Lett. {\bf{103}}, 045002 (2009).


\bibitem{sargis2005-01}S.~Ter-Avetisyan et al., J.\ Phys.~D -- Applied Physics {\bf 38}, 863 (2005).

\bibitem{Krushelnick2005-01} K.~Krushelnick et al., Plasma Physics and Controlled Fusion {\bf{47}}, B451 (2005).
\bibitem{nuernberg2009} F.~Nuernberg et al., Review of Scientific Instruments {\bf{80}}, 033301 (2009).

\bibitem{Esirkepov2006} T.~Esirkepov, M.~Yamagiwa, and T.~Tajima, Phys.\ Rev.\ Lett.\ {\bf 96}, 105001 (2006).

\bibitem{fuchs2006} J.~Fuchs et al., Nature Phys. {\bf2}, 48 (2006).

\bibitem{schreiber2006-01} J.~Schreiber, Phys.\ Rev.\ Lett.\ {\bf 97}, 045005 (2006).

\bibitem{Andreev2008-01}A.\,A.~Andreev et al., Phys.\ Rev.\ Lett. {\bf 101}, 155002
(2008).

\bibitem{Vshivkov1998-01} V.\,A.~Vshivkov et al., Phys.\ Plasmas {\bf 5}, 2727 (1998).

\bibitem{mora2003} P.~Mora, Phys.\ Rev.\ Lett.\ {\bf 90}, 185002 (2003); Phys.\
Rev. E {\bf 72}, 056401 (2005).

\bibitem{Mako1984}F.~Mako and T.~Tajima, Phys. of Fluids {\bf 27}, 1815 (1984).

\bibitem{yan2009}X.\,Q.~Yan et al., Appl. Phys. B, (2009), accepted.

\bibitem{nickles2007-01} P. V.~Nickles, Laser and Particle Beams\ {\bf 25} , 347 (2007).






\end{thebibliography}
\end{document}